\documentclass[]{article}

\AtBeginDocument{%
  }

 \usepackage{amsmath}
 \usepackage{amssymb}
\usepackage{stmaryrd}
\usepackage{dutchcal}
\usepackage{mathtools, nccmath}
\usepackage{lipsum}
\usepackage{fancyvrb}
\usepackage{listings}
\usepackage{graphicx}
\usepackage{lipsum}
\usepackage{multirow}
\usepackage{xcolor}
\usepackage{tcolorbox}
\tcbuselibrary{listings, skins}
\usepackage{array}
\usepackage{booktabs}
\usepackage{tikz}
\usetikzlibrary{positioning, fit}
\usepackage{tabularx}
\usepackage{fitbox}
\usepackage{fullpage}
\usepackage{times}

\lstdefinelanguage{MPGQ}{
    morekeywords={
        MATCH, WHERE, AND, NOT, OR, RETURN, AS, KEY, VAL, ELEMENTOF, SUBSETEQ, LABEL 
    },
    sensitive=True, 
    morecomment=[l]{//}, 
    morecomment=[s]{/*}{*/}, 
    morestring=[b]", 
}

\definecolor{lightgray}{gray}{0.95}
\definecolor{teal}{RGB}{0,128,128}
\definecolor{lightteal}{RGB}{176,224,230}
\definecolor{lightblue}{RGB}{173,216,230}
\definecolor{lightyellow}{RGB}{255,255,224}

\newtcblisting{query}[3][]{
  listing only,
  listing options={language=#3},
  colback=lightgray,
  colframe=teal,
  boxrule=0pt,
  title=#2,
  fonttitle=\bfseries,
  #1
}

\newcommand{\singlebox}[2]{
    \begin{tcolorbox}[enhanced, colback=lightgray, colframe=teal, title={#1}, width=\columnwidth, boxrule=0pt,  fonttitle=\bfseries]
        \begin{center}
        #2
        \end{center}
        \end{tcolorbox}
}

\newcommand{\singleboxs}[2]{
    \begin{tcolorbox}[enhanced, colback=lightgray, colframe=olive, title={#1}, width=\columnwidth, boxrule=0pt,  fonttitle=\bfseries]
        \begin{center}
        #2
        \end{center}
    \end{tcolorbox}
}

\newcommand{\noop}[1]{}

\newcommand{\MetaPG}{Meta-Property Graph}
\newcommand{\MetaPGs}{Meta-Property Graphs}
\newcommand{\metaPG}{meta-property graph}

\newcommand{\MPG}{MPG}
\newcommand{\MPGs}{MPGs}
\newcommand{\MGPML}{MetaGPML}

\newcommand{\substructure}{sub-structure}
\newcommand{\IDs}{\ensuremath{\mathcal{I}}}
\newcommand{\olabel}{\ensuremath{\ell}}
\newcommand{\olabels}{\ensuremath{\mathcal{L}}}
\newcommand{\key}[1]{\ensuremath{Key(#1)}}
\newcommand{\pkey}{\ensuremath{\mathcal{k}}}
\newcommand{\pkeys}{\ensuremath{\mathcal{K}}}
\newcommand{\val}[1]{\ensuremath{Val(#1)}}
\newcommand{\values}{\ensuremath{\mathbb{V}}}
\newcommand{\pvalue}{\ensuremath{\mathcal{v}}}
\newcommand{\pvalues}{\ensuremath{\mathcal{V}}}

\newcommand{\labels}{\ensuremath{L}}
\newcommand{\compatible}{\ensuremath{\mathbb{C}}}
\newcommand{\variables}{\ensuremath{\mathcal{X}}}
\newcommand{\expression}{\ensuremath{\mathcal{E}}}
\newcommand{\clauses}{\ensuremath{\mathcal{C}}}
\newcommand{\queries}{\ensuremath{\mathcal{Q}}}
\newcommand{\pattern}{\ensuremath{\pi}}
\newcommand{\graphPatterns}{\ensuremath{\Pi}}
\newcommand{\conditions}{\ensuremath{\Phi}}

\newcommand{\constant}[1]{\texttt{#1}}
\newcommand{\syntaxFont}[1]{\text{\ttfamily{#1}}}

\newtheorem{theorem}{Theorem}[section]

\newtheorem{definition}[theorem]{Definition}

\begin{document}

\title{Meta-Property Graphs: Extending Property Graphs with Metadata Awareness and Reification}

\author{Sepehr Sadoughi, Nikolay Yakovets, George Fletcher}


\date{Eindhoven University of Technology, The Netherlands \\  \texttt{$\{$s.sadoughi, n.yakovets, g.h.l.fletcher$\}$@tue.nl}}

\maketitle
\begin{abstract}
The ISO standard Property Graph model has become increasingly popular for representing complex, interconnected data. However, it lacks native support for querying metadata and reification, which limits its abilities to deal with the demands of modern applications. We introduce the vision of \MetaPG{}, a backwards compatible extension of the property graph model addressing these limitations. Our approach enables first-class treatment of labels and properties as queryable objects and supports reification of substructures in a graph. We propose \MGPML{}, a backwards compatible extension of the Graph Pattern Matching Language forming the core of the ISO standard GQL, to query these enhanced graphs. We demonstrate how these foundations pave the way for advanced data analytics and governance tasks that are challenging or impossible with current property graph systems.
\end{abstract}
\section{Introduction}
\label{sec:introduction}
Modern data engineering applications increasingly demands flexibility and agility in handling data and metadata more than ever before.
During exploratory analytics, structure emerges gradually without a predefined schema. Heterogeneity in what is a data value versus what is an attribute name is inherent during discovery, profiling, and exploration.
In data sources integration scenarios, such as building knowledge graphs, data and schema heterogeneity is inevitable:
what appears as a data value in one source might be a node label or property name in another,
while a subgraph in one source might correspond to a single node elsewhere.
Modern data management solutions must therefore fully support heterogeneity and fluid boundaries between data and metadata.

Metadata is commonly understood in two distinct ways. The first is {\em attribute metadata}- characteristics associated with data objects, such as column names or table names in relational databases or node labels in graphs. For instance, when storing contact information, attributes like name and email are metadata.
The second form is {\em reification}- representing an aggregation of complex relationships or structures into a new entity or data object, making it easier to manage and analyze. In the context of entity-relationship modeling, reification is often used to transform relationship sets into entity sets, allowing for more effective data modeling and manipulation. For example, in an e-commerce system, reifying the relationship between customers and orders creates an order history entity that can be analyzed and queried.\footnote{Other more complex types of metadata, such as reflective or active data \cite{stijn,Fletcher14,divesh}, are beyond the scope of our current discussion.}


The ISO standard Property Graph (PGs) model has gained popularity in graph data management and is widely adopted, e.g., in graph DB systems such as Neo4j, Tigergraph, and Amazon Neptune, as well as relational DB systems such as DuckDB which implement the ISO extensions to SQL for PG querying. 
In a  property graph nodes and edges are labeled and also have associated sets of property name/value pairs (e.g., a node labeled \texttt{Person} with property \texttt{Birthdate} having data value 11-11-2001; here \texttt{Person} and \texttt{Birthdate}  are attribute metadata).
While PGs offer a model closely aligned with conceptual domain representations, the model makes (1) a strict distinction between metadata and data, and (2) has no support for reification. In contrast, the W3C's Resource Description Framework (RDF) for modeling graph data treats everything as data, including attributes of nodes and edges, and natively supports reification. This capability makes RDF particularly useful for applications requiring sophisticated data validation, verification, querying, and governance of metadata. There is an opportunity to bring these powerful features to PGs while preserving their core design principles and inherent strengths in handling data heterogeneity.

In this paper we present a vision for overcoming barriers to flexible management of data-metadata heterogeneity in PG data management applications.
Towards this, we introduce \MetaPGs{} (MPG), a fully backwards compatible extension of the PG model that addresses limitations in representing and querying metadata. Our approach enables first-class treatment of labels and properties as queryable objects, as well as reification of subgraphs. On this foundation, we further propose \MGPML{}, a fully backwards compatible extension of the Graph Pattern Matching Language (GPML), the core language at the heart of the ISO standard GQL for PG querying.  We give a complete formal specification of MPG and \MGPML{}, providing the foundations for our vision. Furthermore, we demonstrate how these contributions facilitate advanced data analytics, integration, and governance tasks that are challenging or impossible with current PG systems. 

\section{Related Work}
\label{sec:related_work}
The PG model represents a design point on the continuum between the Relational (fixed metadata) and RDF (no fixed metadata) data models. 
While by default schema-less, supporting agile modeling and development, PGs nonetheless make a strict distinction between data values and metadata values.
Of course, all points along this design continuum are equally important, finding their applications and use-cases.   And indeed, data cleaning, wrangling, integration, and exchange  are very often about moving (meta)data along the continuum. 
What has been missing is an appropriate design for PGs to fully meet modeling and querying needs of data management scenarios such as these 
(see Chapter 2 of \cite{bonifati} for a survey of design approaches for PG modeling).


In the research literature, dealing with the challenges of data-metadata heterogeneity was studied in the context of relational data integration and data integration on the web, leading to 
solutions for relational and XML data-metadata mapping and exchange (e.g.,
\cite{BonifatiCHLPC10,fletcher_towards_2009,HernandezPT08}) and 
seminal work such as the SchemaSQL and FISQL data-metadata query languages for relational databases
\cite{lakshmanan_schemasql_2001,wyss_relational_2005}.  Our work approaches these challenges in the new context of the ISO standards for graph data management.

In another direction, there is a growing body of work on mappings between graph data models \cite{mappings_4,mappings_1,mappings_2,mappings_3,StoicaFS19} and generalizing both PG and RDF \cite{gelling_bridging_2023,onegraph,multilayer_graphs}.
Our vision is orthogonal and complementary to these investigations, aiming at extending the PG model and the standardized GQL and SQL/PGQ query languages with seamless data-metadata functionalities.

\section{A guided tour of working with \MPGs{}}
\label{sec:metaPG}

\begin{figure*}
  \centering
  \includegraphics[width=\textwidth]{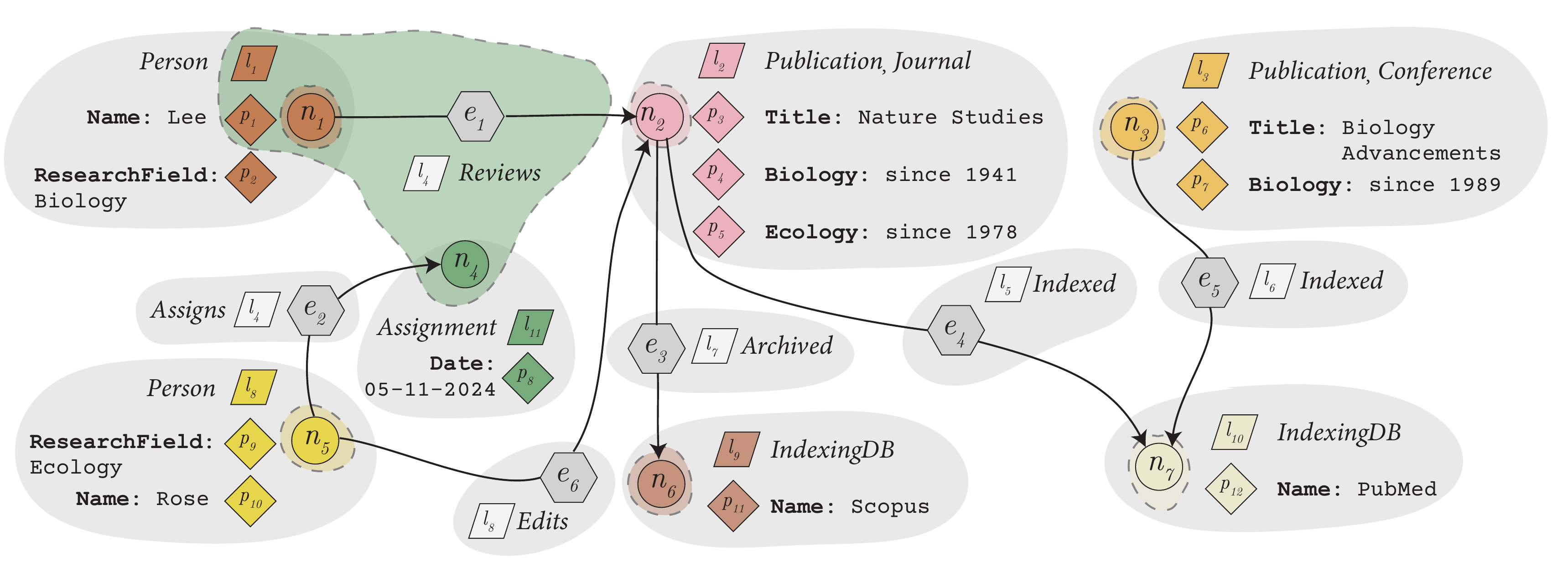}
  \caption{Example \metaPG{} \textbf{$G$}}
  \label{fig:MPG_figure}
\end{figure*}

%
Figure \ref{fig:MPG_figure} illustrates a simple MPG, demonstrating how the MPG model enables direct querying of metadata as data objects. The sample database represents publications, indexing databases, and persons with their relationships, modeled using four types of data objects: edges ($E_G$), nodes ($N_G$), properties ($P_G$), and labels ($L_G$). Note that properties and labels are now first class citizens, unlike in the current PG standard. Table 1 introduces the pattern notation used in our examples, which will be formally defined in Section \ref{sec:MetaGPML}.

\begin{tcolorbox}[
    colback=lightgray,
    colframe=olive,
    boxrule=0pt,
    width=\columnwidth,
    title={Table 1: Data objects pattern notation},
    fonttitle=\bfseries,
    enhanced,
    left=0pt, right=0pt, top=0pt, bottom=0pt,
    toptitle=0pt, bottomtitle=0pt,
    lefttitle=0pt, righttitle=0pt
]
{\small
\begin{tcolorbox}[
    colback=lightgray,
    colframe=gray!20,
    title={\textcolor{olive}{\textbf{Nodes}}},
    fonttitle=\bfseries,
    boxrule=0pt,
    width=\columnwidth,
    left=0pt, right=0pt, top=0pt, bottom=0pt,
    toptitle=0pt, bottomtitle=0pt,
    lefttitle=0pt, righttitle=0pt
]
    \begin{tabular}{@{}p{3cm} p{0.65\columnwidth}@{}}
        \syntaxFont{(x:l)} & node variable (nv.) \syntaxFont{x} with label \syntaxFont{l} \\
        \syntaxFont{(x:l).z} & nv. \syntaxFont{x} with label \syntaxFont{l} and property variable \syntaxFont{z} \\
        \syntaxFont{(x::$\pi$)} & nv. \syntaxFont{x} with a pattern $\pi$ in it's reified substructure
    \end{tabular}
\end{tcolorbox}

\begin{tcolorbox}[
    colback=lightgray,
    colframe=gray!20,
    title={\textcolor{olive}{\textbf{Edges}}},
    fonttitle=\bfseries,
    boxrule=0pt,
    width=\columnwidth,
    left=0pt, right=0pt, top=0pt, bottom=0pt,
    toptitle=0pt, bottomtitle=0pt,
    lefttitle=0pt, righttitle=0pt
]
    \begin{tabular}{@{}p{3cm} p{0.65\columnwidth}@{}}
        \syntaxFont{-[x:l]->} & edge variable \syntaxFont{x} with label \syntaxFont{l} \\
        \syntaxFont{-[x:?y]->} & edge variable \syntaxFont{x} with label set variable \syntaxFont{y} \\
        \syntaxFont{-[x].z->} & edge variable \syntaxFont{x} with property variable \syntaxFont{z}
    \end{tabular}
\end{tcolorbox}

\begin{tcolorbox}[
    colback=lightgray,
    colframe=gray!20,
    title={\textcolor{olive}{\textbf{Properties}}},
    fonttitle=\bfseries,
    boxrule=0pt,
    width=\columnwidth,
    left=0pt, right=0pt, top=0pt, bottom=0pt,
    toptitle=0pt, bottomtitle=0pt,
    lefttitle=0pt, righttitle=0pt
]
    \begin{tabular}{@{}p{3cm} p{0.65\columnwidth}@{}}
        \syntaxFont{\{x\}} & property variable \syntaxFont{x} \\
        \syntaxFont{KEY(x)}, \syntaxFont{VAL(x)} & key and value of property variable \syntaxFont{x}
    \end{tabular}
\end{tcolorbox}

\begin{tcolorbox}[
    colback=lightgray,
    colframe=gray!20,
    title={\textcolor{olive}{\textbf{Labels}}},
    fonttitle=\bfseries,
    boxrule=0pt,
    width=\columnwidth,
    left=0pt, right=0pt, top=0pt, bottom=0pt,
    toptitle=0pt, bottomtitle=0pt,
    lefttitle=0pt, righttitle=0pt
]
    \begin{tabular}{@{}p{3cm} p{0.65\columnwidth}@{}}
        \syntaxFont{|x|} & label set variable \syntaxFont{x} \\
        \syntaxFont{c ELEMENTOF x} & check if label \syntaxFont{c} exists in label set variable \syntaxFont{x}
    \end{tabular}
\end{tcolorbox}
}
\end{tcolorbox}

\subsection{Working with Data, Labels, and Properties}
\label{sec:data_metadata_interaction}

\MetaPG{} enables us to query directly on label sets as recognized data objects without considering the nodes or edges that they are associated with. This will facilitate analytics over classes in the database and work with this data independently. 
As a simple instance, in the following example \MGPML{} query $Q_1$, we retrieve the label set that contains labels \constant{Publication} in the database to figure out the other co-occurring tags regardless of which node they belong to.    

\begin{query}{$Q_1$: Which label sets contain the label \constant{Publication}?}{MPGQ}   
MATCH |l|
WHERE "Publication" ELEMENTOF l
RETURN l AS "Publication_Co_Tags"
\end{query}

\singlebox{Result table of $Q_1$}{%
\begin{tabular}{|l|}
    \hline
  \textbf{Publication\_Co\_Tags} \\ \hline
  \textcolor{black}{\text{\{"Publication", "Journal"\}}} \\
  \textcolor{black}{\text{\{"Publication", "Conference"\}}}  \\
    \hline
\end{tabular}
}

Here \texttt{l} is a variable which binds to label set objects.
With \MGPML{} we can also query properties as independent data objects. Query $Q_2$ demonstrates this by retrieving the values of all \constant{Name} properties, regardless of whether the property is attached to a node or an edge. 

\begin{query}{$Q_2$: What are the values of \constant{Name} properties?}{MPGQ}
    MATCH {p} 
    WHERE KEY(p) = "Name"
    RETURN VAL(p) AS "Names"
\end{query}

\singlebox{Result table of $Q_2$}{
    \small
    \begin{tabular}{|c|}
        \hline
            \textbf{Names} \\ 
            \hline
            Lee \\
            Scopus \\
            Rose \\  
            PubMed \\ 
            \hline
    \end{tabular}
}

In general, this capability of MetaPGML to bind property names and label sets to variables allows the seamless promotion of data to metadata, and vice versa, as is critical in applications such as data integration, cleaning, and governance. The $Q_3$ and $Q_4$ examples demonstrate such \textit{promotions} and \textit{demotions} of (meta)data.  The following example aims to match the relationships between publications and the indexing databases and return the results in a structure to show in which indexing database each publication is being archived or indexed. 

\begin{query}{$Q_3$: Which relationships does each publication have with each indexing databases?}{MPGQ}
MATCH (x:Publication)-[:?y]->(z:Indexing_DB)
RETURN x.Title AS "Title", LABEL(y) AS
z.Name
\end{query}

\noindent Query results are shown below, where \syntaxFont{LABEL} resolves label sets bound to \syntaxFont{y}, while \syntaxFont{z.Name} values form column headers: (for convenience we switch between tabular and binding set representations of query results)

\singlebox{Result bindings of $Q_3$}{%
\begin{equation*}
    \begin{aligned}[b]
        \{
    &\left( \ \text{Title} \mapsto \textcolor{teal}{\text{"Nature Studies"}},  \text{Scopus} \mapsto \textcolor{teal}{\text{\{"Archived"\}}}  \ \right),\\
    &\left( \ \text{Title} \mapsto \textcolor{teal}{\text{"Nature Studies"}},  \text{PubMed} \mapsto \textcolor{teal}{\text{\{"Indexed"\}}}  \ \right),\\
    &\left( \ \text{Title} \mapsto \textcolor{teal}{\text{"Biology Advancements"}},  \text{PubMed} \mapsto \textcolor{teal}{\text{\{"Indexed"\}}}  \ \right)
        \} \\
    \end{aligned}
\end{equation*}}

Our next query ($Q_4$) demonstrates how the \MGPML{} can manipulate metadata as regular data.
This query finds potential reviewers for publications based on their research fields and the publication's acceptance criteria, showcasing how MPG treats metadata (property keys) as queryable data, enabling more flexible and powerful queries than the PG model.

\begin{query}{$Q_4$: Who are potential reviewer candidates for each publication based on the candidate's research fields?}{MPGQ}
MATCH (x:Person), (y:Publication).z
WHERE x.ResearchField = KEY(z)
RETURN x.Name AS "Reviewer candidate", 
       y.Name AS "Publication venue",
       KEY(z) AS "Research field"
\end{query}
\singlebox{Result table of $Q_4$}{%
    \begin{tabular}{|c|c|c|}
        \hline
        \textbf{Reviewer candidate} & \textbf{Publication venue} & \textbf{Research field}  \\
        \hline
        Lee & Nature Studies & Biology  \\
        Lee & Biology advancement & Biology  \\
        Rose & Nature Studies  & Ecology \\
        \hline
\end{tabular}
}

The structure of the graph requires demoting the metadata of the property key in \constant{Publication}, indicating when it started publishing in a certain research field. This allows comparison with the \constant{ResearchField} property of \constant{Person} entities, enabling identification of reviewer candidates for different publications.

\subsection{Working with Reification}
\label{sec:reification}

\MPG{} and \MGPML{} enable reification as the second type of metadata on property graphs. A \substructure{} of an MPG can be reified as a node, thereby making it a first-class citizen. This enables annotating part of the graph by assigning properties and labels to a node that reifies that part. It is worth mentioning that the part of the graph that is being reified may or may not be a proper graph, which means that a node can reify some data objects without their assigned relationships in the graph, e.g., a set of specific properties or relationships, can be reified without the nodes they are associated with.   

In Figure \ref{fig:MPG_figure}, we represent the statement ``Rose assigned Lee as a reviewer on 5th November 2024'' using node $n_4$ that induces a \substructure{} containing node $n_1$, some relevant properties, and outgoing \constant{review} edges. Query $Q_5$ retrieves the name of the person who assigns a reviewer:

\begin{query}{$Q_5$: Who assigned Lee as a reviewer and when?}{MPGQ}   
MATCH  (x:Person)-[:assigns]->
       (y::(z:Person)-[:reviews]->())
WHERE  z.Name = "Lee"
RETURN  z.Name AS "reviewer name",
        y.Date AS "Date",
        x.Name AS "Assigning editor"
\end{query}
\noindent Note that in the \syntaxFont{MATCH} clause we have a graph pattern embedded in a node pattern, to denote a query to be executed on the sub-structure reified by the node which is bound to variable \syntaxFont{y}.
\singlebox{Result table of $Q_5$}{%
\begin{tabular}{|c|c|c|}
    \hline
     \textbf{Reviewer name} & \textbf{Date} & \textbf{Assigning editor}\\ \hline
     Lee &  \ 05-11-2024 \ & Rose \\
     \hline
\end{tabular}
}

While it may look that MPG is a hypergraph model it's important to note that hypergraphs focus on complex relationships, while reification focuses on complex objects. In MPG, edges still connect exactly two nodes. Nodes can act as meta-nodes, with other data objects (nodes, edges, properties, label sets) assigned for reification.

\section{\MetaPG{} Model}
\label{sec:metaPG_model}

We next formalize \MetaPGs.  Let \IDs, \olabels, \pkeys, and \pvalues{} be pairwise disjoint sets of object identifiers, labels, property keys, and property values, respectively.

\begin{definition} \label{def:mpg}
    A \metaPG{} is a directed and undirected vertex- and edge-labeled graph \break \(G=(N,E,P,L,\lambda,\mu,\sigma, \upsilon,\eta,\rho)\), where:
    \begin{itemize}
        \setlength\itemsep{0.3em}
        \item \(N, E, P, L \subseteq \IDs\) are finite, pairwise disjoint sets,
        \item \(\mu : L \rightarrow 2^\olabels\) assigns a finite set of labels to each label identifier,
        \item \(\lambda : N\cup E \rightarrow L\) is a bijective labeling function assigning a label set identifier to each node and edge,
        \item \(\upsilon : P \rightarrow \pkeys \times \pvalues\) assigns key-value pairs to properties,
        \item \(\sigma : N\cup E \rightarrow \compatible\) assigns \textit{compatible} property sets to nodes and edges, such that
       for each pair of distinct objects \( o_1, o_2 \in N \cup E \), it holds that \( \sigma{(o_1)} \cap \sigma{(o_2)} = \varnothing \) and \( \bigcup_{o\in N\cup E} \sigma(o)  = P \), i.e., every and each property \(p \in P\) is assigned to exactly one node or edge.
        
        \item \(\eta = (\eta_s, \eta_t, \eta_u)\) where:
            \begin{itemize}
                \item \(\eta_s, \eta_t : E_d \rightarrow N\) assign source and target nodes to directed edges,
                \item \(\eta_u : E_u \rightarrow \{\{u,v\} \mid u,v \in N\}\) assigns node pairs to undirected edges,
            \end{itemize}
        \item \(\rho : N \rightarrow 2^{N\cup E\cup P\cup L}\) assigns finite sets of objects to nodes, 
        such that for each $n \in N$, it holds that $n \notin \rho^*(n)$, where $\rho^*(n)$ is the closure of $\rho(n)$,\footnote{Formally, 
$\rho^0(n) = \rho(n)$, 
$\rho^i(n) = \rho^{i-1}(n) \cup \bigcup_{n'\in N\cap \rho^{i-1}(n)} \rho(n')$, and
$\rho^*(n) = \bigcup_{i=0}^\infty \rho^i(n)$.} 
ensuring that the \substructure{} associated with each node is well-founded.
\end{itemize}
    where \(E = E_d \cup E_u\) the set of compatible property sets \(\compatible = \{ C \subseteq P \mid \forall p_1,p_2 \in C, p_1 \neq p_2 \Rightarrow \key{p_1} \neq \key{p_2}\}\),
    \(\key{p} = \pi_1(\upsilon(p))\), \(\val{p} = \pi_2(\upsilon(p))\).
\end{definition}

\paragraph*{Design decisions.}
The constraints we have imposed in our design of MPGs are very minimalistic, ensuring only that all properties are uniquely assigned to edges and nodes, and that reification is well-founded. We have deliberately avoided imposing unwarranted constraints. For instance, there is no constraint preventing a node from having an outgoing edge to a subgraph containing one of the node's own properties. This flexibility allows for representing scenarios such as ``Mary legally changed her first name'' or ``Mary audits all names in the database (including her own)''. Similarly, there is no constraint requiring that for each edge in $\rho(n)$, the source and target of the edge must also be in $\rho(n)$, which, for instance, can be particularly valuable in applications with privacy preservation concerns. As an example, This permits representations like ``Mary is confident that Bob edited a book, but not confident about which book'', where only the node representing Bob and Bob's outgoing edge labeled 'edited' are included, without the target of this edge. Of course, such additional constraints can be added (or existing constraints can be removed) as appropriate to the specific application domain.

It is worth noting that MPGs support meta-properties, i.e., properties on properties. For instance, we can represent ``John's birthdate was entered on 23 March 2020'' since $\rho$ can consist of just a singleton containing a property. For simplicity, we have omitted special treatment of meta-properties in our main presentation.

We next formally define the graph substructure associated with a node.

\begin{definition} (Sub-Structure) 
Let $G=(N,E,P,L,\lambda,\mu,\sigma,\upsilon,\eta,\rho)$ be a \metaPG{} and $n\in N$.
The \substructure{} of $G$ induced by $n$ is $G_n=(N_n, E_n, P_n, L_n, \lambda_n, \mu_n, \sigma_n, \upsilon_n, {\eta}_n, \rho_n)$ where:
\begin{itemize}
    \setlength\itemsep{0.3em}
    \item $N_n = N \cap \rho(n)$, $E_n = E \cap \rho(n)$, $L_n = L \cap \rho(n)$, $P_n = P \cap \rho(n)$
    \item $\upsilon_n = \upsilon|_{P_n}$, $\mu_n = \mu|_{L_n}$ (domain restrictions)
    \item $\lambda_n(o) = \lambda(o)$ for $o \in N_n \cup E_n$ if $\lambda(o) \in \rho(n)$, undefined otherwise
    \item $\sigma_n(o) = \sigma(o) \cap \rho(n)$ for $o \in N_n \cup E_n$
    \item ${E_d}_n, {E_u}_n \subseteq E_n$ are directed and undirected edge subsets
    \item \({\eta}_n = ({\eta_s}_n, {\eta_t}_n, {\eta_u}_n)\) where:
    \begin{itemize}
        \item For $e \in {E_d}_n$:
            \begin{itemize}
                \item ${\eta_s}_n(e) = \eta_s(e)$ if $\eta_s(e) \in \rho(n)$, undefined otherwise
                \item ${\eta_t}_n(e) = \eta_t(e)$ if $\eta_t(e) \in \rho(n)$, undefined otherwise
            \end{itemize}
        \item ${\eta_u}_n(e) = \eta_u(e) \cap \rho(n)$ for $e \in {E_u}_n$
    \end{itemize}
    \item $\rho_n(n') = \rho(n) \cap \rho(n')$ for $n' \in N_n$, satisfying the same well-foundedness property as in Definition \ref{def:mpg}. 
\end{itemize}
\end{definition}

Intuitively, the \substructure{} $G_n$ can be thought of as a "slice" or "view" of the original \metaPG{} $G$, focused on the node $n$ and its associated elements. This substructure includes only those nodes, edges, properties, and labels that are explicitly linked to $n$ through the reification function $\rho$. It maintains the original graph's structure and relationships, but only for the elements within this subset. This allows us to zoom in on specific parts of the graph, making it easier to analyze and query complex, nested structures without losing the context of how they fit into the larger graph.

\section{\MetaPG{} Pattern Matching Language} 
\label{sec:MetaGPML}

 We next formalize our Graph Pattern Matching Language for Metaproperty Graphs (\MGPML{}) as an extension to the Graph Pattern Matching Language (GPML) \cite{deutsch_graph_2022,francis_gpc_2023,francis_researchers_2023}. GPML is the core element of both SQL/PGQ and GQL languages standardized by ISO. By extending GPML we ensure backward compatibility in the sense that every GPML query is a \MGPML{} query. 

\subsection{Syntax} 
\label{sub-sec:syntax}

 Let \variables{} be a countable infinite set of variables. We introduce descriptors \(\delta_1\), \(\delta_2\), and \(\delta_3\) in \MGPML{}, consistent with GPML \cite{francis_gpc_2023}. For $ \syntaxFont{x}, \syntaxFont{y}, \syntaxFont{z}\in \variables, \ \olabel \in \olabels, \ \pkey \in \pkeys, $ and $ \pvalue \in \pvalues $, we have
\singleboxs{Descriptors:}{
\begin{equation} \label{expressions:descriptor}
    \begin{aligned}
        \delta_{nodes} &:= \delta_{edges} \mid \syntaxFont{::}\pi \mid \syntaxFont{:} \olabel \syntaxFont{::} \pi \mid \syntaxFont{x::}\pi \mid \syntaxFont{x:} \olabel \syntaxFont{::} \pi \\
        &\quad \mid \syntaxFont{:?y::}\pi \mid \syntaxFont{x:?y::} \pi \\
        \delta_{edges} &:= \syntaxFont{x} \mid \syntaxFont{:} \olabel \mid \syntaxFont{:?y} \mid \syntaxFont{x:} \olabel \mid \syntaxFont{x:?y} \\
    \end{aligned}
\end{equation}}

\noindent The descriptors in (\ref{expressions:descriptor}) will be used in the following patterns to assign variables and constants to different data objects. A pattern $\pi$ in \MGPML{} and operations on it can be formulated as follows: 

\singleboxs{Patterns:}{
\small
\begin{eqnarray} \label{expressions:patterns}
    \begin{array}{llr}
        \pattern  := & \syntaxFont{(} \delta_{nodes} \syntaxFont{)} \ \mid \ \syntaxFont{()} \ & \text{(node pattern)}
        \\
        &\mid \ \syntaxFont{(} \delta_{nodes} \syntaxFont{).x} \ \mid \ \syntaxFont{().x} & \\ 
        & \mid \ \syntaxFont{-[} \delta_{edges} \syntaxFont{]->} \ \mid \ \syntaxFont{<-[} \delta_{edges} \syntaxFont{]-} & \text{(edge pattern)}\\
        &\mid \syntaxFont{-[} \delta_{edges} \syntaxFont{]-}&\\
        & \mid \ \syntaxFont{-[]->} \ \mid \ \syntaxFont{<-[]-} \ \mid \syntaxFont{-[]-} & \\
        & \mid \ \syntaxFont{-[} \delta_{edges} \syntaxFont{].x} \syntaxFont{->} \ \mid \ \syntaxFont{<-[} \delta_{edges} \syntaxFont{].x} \syntaxFont{-}  & \\
        & \mid \ \syntaxFont{-[} \delta_{edges} \syntaxFont{].x} \syntaxFont{-} & \\
        & \mid \ \syntaxFont{\{x\}} \ \mid \ \syntaxFont{\{\}} & \text{(property pattern)} \\
        & \mid \ \ \syntaxFont{|x|} \ \mid \ \syntaxFont{||}& \text{(label pattern)} \\
        & \mid \ \pi\pi & \text{(concatenation)} \\
        & \mid \ \pi \syntaxFont{+} \pi & \text{(union)} \\
        & \mid \ \pi \ \syntaxFont{WHERE} \ \Phi & \text{(conditioning)} \\
        \graphPatterns := & \pi \mid \Pi, \Pi & \text{(graph pattern)} \\
    \end{array}
\end{eqnarray}}

\noindent  We define $\expression$ as the \textit{expression} for \textit{condition} $\Phi$ that can be used in a pattern after the \syntaxFont{WHERE} keyword as follows, considering
$ c \in \pkeys \cup \pvalues \cup \olabels$.
\singleboxs{Expressions and conditions:}{
\begin{eqnarray} \label{expressions:conditions}
    \begin{array}{llr}
        \expression := & \ \syntaxFont{x} \mid \ \syntaxFont{x}.\pkey  \ \mid \ \pvalue \ \mid \ \pkey \ \mid \ \olabel & \multirow{2}{*}{\text{(expressions)}}\\
        & \ \mid \ \syntaxFont{KEY}(\syntaxFont{x}) \ \mid \ \syntaxFont{VAL}(\syntaxFont{x}) \mid \syntaxFont{LABEL}(\syntaxFont{x}) &\\
        
        \Phi  := & \expression = \expression \ \mid \ \expression < \expression \ \mid \ \syntaxFont{x}:\olabel & \multirow{3}{*}{\text{(conditions)}} \\ 
        &\mid \syntaxFont{SUBSETEQ}(\syntaxFont{x}, \ \syntaxFont{y}) \ \mid \  c \ \syntaxFont{ELEMENTOF} \ \syntaxFont{y} & \\
        &\mid \ \Phi \ \syntaxFont{AND} \ \Phi \ \mid \ \Phi \ \syntaxFont{OR} \ \Phi \ \mid \ \syntaxFont{NOT} \ \Phi & 
    \end{array}
\end{eqnarray}}

It is important to note that $\syntaxFont{KEY}(\syntaxFont{x})$ and $\syntaxFont{VAL}(\syntaxFont{x})$ operate on properties as first-class data objects, similar to nodes and edges. The same applies to $\syntaxFont{LABEL}(\syntaxFont{x})$, which operates on label sets as distinguishable data objects and retrieves the set of label strings it contains. These functions should not be confused with similar functions in some other PG query languages like OpenCypher, which only simulates these capabilities at the query language level within the conventional PG data model. Finally, based on (\ref{expressions:patterns}) and (\ref{expressions:conditions}), we formulate the definition of \textit{clauses} and \textit{queries}.

\singleboxs{Clauses and queries:}{
\begin{eqnarray} \label{expressions:queries}
    \begin{array}{ll}
        \clauses := & \syntaxFont{MATCH} \ \graphPatterns  \\
        & \ \mid \ \syntaxFont{FILTER} \ \conditions  \\
        
        \queries  := & \clauses \queries  \\ 
        &\mid \ \syntaxFont{RETURN} \ \expression_1 \ \syntaxFont{AS} \ x_1, \ ... \ ,\expression_n \ \syntaxFont{AS} \ x_n \\
    \end{array}
\end{eqnarray}}

The syntax of \MGPML{} is designed to be intuitive and expressive, building upon familiar concepts from the GQL standard query language. Node and edge patterns use parentheses and brackets respectively, with labels prefixed by colons. The double-colon syntax (\syntaxFont{::}) introduces subgraph patterns, enabling querying of reified structures. 
To distinguish them from constants, variables are prefixed with question marks when used to bind labels or properties. The language includes specialized predicates like \syntaxFont{ELEMENTOF} and \syntaxFont{SUBSETEQ} for working with sets of labels and properties.


\subsection{Semantics}
\label{sub-sec:semantics}

Next, we present the semantics of our language as a clear, unambiguous foundation for further study and implementation. To avoid unnecessary complexity, we presented the most vital semantics to illustrate the concept to understand how \MGPML{} works. The complete and more detailed semantics covering all situations are presented in appendix \ref{sec:appendixA}.

\begin{definition} (Bindings)
    A binding $\beta: \variables \rightarrow \values$ assigns variables $x \in \variables$ to values $v \in \values$, where $\values = \IDs \cup \olabels \cup \pkeys \cup \pvalues$. We denote this as $(x_1 \mapsto v_1, ..., x_n \mapsto v_n)$, where $x_1, ..., x_n$ are the variables in the domain of $\beta$ $(\text{Dom}(\beta))$ and $v_1, ..., v_n$ are their corresponding values. We denote the empty binding as $()\,$.
\end{definition}

\begin{definition} (Compatibility of Bindings and their Join)
Two bindings $\beta_1, \beta_2$ are compatible, denoted by $\beta_1 \sim \beta_2$, if they agree on their shared variables. Specifically, for every $x \in \text{Dom}(\beta_1) \cap \text{Dom}(\beta_2)$, it holds that $\beta_1(x) = \beta_2(x)$.

When $\beta_1 \sim \beta_2$, their join $\beta_1 \Join \beta_2$ is defined as follows: $\text{Dom}(\beta_1 \Join \beta_2) = \text{Dom}(\beta_1) \cup \text{Dom}(\beta_2)$. For any variable $x$, $(\beta_1 \Join \beta_2)(x) = \beta_1(x)$ if $x \in \text{Dom}(\beta_1) \setminus \text{Dom}(\beta_2)$, and $(\beta_1 \Join \beta_2)(x) = \beta_2(x)$ if $x \in \text{Dom}(\beta_2)$.
\end{definition}

A pattern matching semantic $\llbracket \pi \rrbracket_{G}$ in \metaPG{} is a set of bindings $\beta$ that assign variables to values.

\singleboxs{Node pattern matching:}{
    \begin{equation}\label{Semantics:node}
    \llbracket \syntaxFont{()} \rrbracket_{G} = \{ () \mid n \in N \}
    \end{equation}

    \begin{equation}\label{Semantics:nodeWithVariable}
    \llbracket \syntaxFont{(x)} \rrbracket_{G} = \{ (\syntaxFont{x} \mapsto n) \mid n \in N \}
    \end{equation}

    \begin{equation}\label{Semantics:nodeWithLabel}
    \llbracket \syntaxFont{(x:\olabel)} \rrbracket_{G} = \{ (\syntaxFont{x} \mapsto n) \mid n \in N, \olabel \in \mu(\lambda(n)) \}
    \end{equation}

    \begin{equation}\label{Semantics:nodeWithLabelVariable}
    \llbracket \syntaxFont{(x:?y)} \rrbracket_{G} = \{ (\syntaxFont{x} \mapsto n, \syntaxFont{y} \mapsto l) \mid n \in N, l = \lambda(n) \}
    \end{equation}
    
    \begin{equation}\label{Semantics:nodeWithNodeAndPropertyVariable}
        \llbracket \syntaxFont{(x).z} \rrbracket_{G}  =  \{ (\syntaxFont{x} \mapsto n, \syntaxFont{z}\mapsto p)  \mid  n \in N , p \in \sigma(n) \}
    \end{equation}
}

Equation \ref{Semantics:node} matches any node $n \in N$. Equation \ref{Semantics:nodeWithVariable} assigns a node to variable \syntaxFont{x}. Equation \ref{Semantics:nodeWithLabel} matches nodes with specific labels, where $\mu(\lambda(n))$ determines the node's labels. Equation \ref{Semantics:nodeWithLabelVariable} assigns a node's label set to variable \syntaxFont{y}. Finally Equation \ref{Semantics:nodeWithNodeAndPropertyVariable} marches nodes with property $p$ assigned to variable \syntaxFont{z}.

Meta-nodes are a key concept in \MGPML{}, defined as nodes to which diverse elements are assigned based on the concept of \substructure{}. This feature enables reification within the \MetaPG, allowing for a more nuanced representation of complex relationships and structures.

Equation \ref{Semantics:meta-nodeWithVariableAndLable} demonstrates the matching of nodes with labels to patterns within their associated \substructure{}:
\singleboxs{Meta-node pattern matching:}{
    \begin{equation}\label{Semantics:meta-nodeWithVariableAndLable}
        \llbracket \syntaxFont{(x:\olabel::} \pi \syntaxFont{)} \rrbracket_{G}  = 
             \llbracket \syntaxFont{(x:\olabel)} \rrbracket_{G } \Join \llbracket \pi \rrbracket_{G_n}
    \end{equation}
}
The meta-node semantic definition involves a full join operation between two patterns divided by \syntaxFont{::} in the node pattern descriptor. Here, $\llbracket \syntaxFont{(x:\olabel)} \rrbracket_{G}$ represents a simple node pattern matching, while $\llbracket \pi \rrbracket_{G_n}$ is a pattern matching within the $G_n$ \substructure{}.
We define the semantics of edges in our model, supporting both directed and undirected edges.

\singleboxs{Edge pattern matching:}{
\small
\begin{align}
    \llbracket \syntaxFont{-\lbrack \rbrack->} \rrbracket_{G} &= \{ ()  \mid  e \in E_d \} \label{Semantics:edge} \\
    \llbracket \syntaxFont{-\lbrack x\rbrack->} \rrbracket_{G} &= \{ (\syntaxFont{x} \mapsto e)  \mid  e \in E_d \} \label{Semantics:edgeWithVariable} \\
    \llbracket \syntaxFont{-\lbrack :\olabel \rbrack->} \rrbracket_{G} &= \{ ()  \mid  e \in E_d , \olabel \in \mu(\lambda(e))\} \label{Semantics:edgeWithLable} \\
    \llbracket \syntaxFont{-\lbrack x:\olabel \rbrack-} \rrbracket_{G} &= \{ (\syntaxFont{x} \mapsto e)  \mid  e \in E_u , \olabel \in \mu(\lambda(e))\} \label{Semantics:udEdgeWithLable} \\
    \llbracket \syntaxFont{-\lbrack x:?y\rbrack->} \rrbracket_{G} &= \{ (\syntaxFont{x} \mapsto e, \syntaxFont{y} \mapsto l)  \mid  e \in E_d, l = \lambda(e)\} \label{Semantics:edgeWithLableVariable} \\
    \llbracket \syntaxFont{-\lbrack x\rbrack.z->} \rrbracket_{G} &= \{ (\syntaxFont{x} \mapsto e, \syntaxFont{z} \mapsto p)  \mid  e \in E_d, p \in \sigma(e) \} \label{Semantics:edgeWithPropertyVariable}
\end{align}
}
Equations \ref{Semantics:edge}-\ref{Semantics:udEdgeWithLable} match various edge patterns, including directed and undirected edges with or without labels. Equation \ref{Semantics:edgeWithLableVariable} allows assigning the edge's label set to variable \syntaxFont{y}, while \ref{Semantics:edgeWithPropertyVariable} enables assigning the edge's properties to variable \syntaxFont{z}.

The \metaPG{} model enables defining specific semantics for property and label objects, enhancing query capabilities on these elements.
\singleboxs{Property and label objects:}{
\begin{align}
    \llbracket \syntaxFont{\{x\}} \rrbracket_{G} &= \{ (\syntaxFont{x} \mapsto p) \mid p \in P \} \label{Semantics:propertyElements} \\
    \llbracket \syntaxFont{|x|} \rrbracket_{G} &= \{ (\syntaxFont{x} \mapsto l) \mid l \in L \} \label{Semantics:labelElements}
\end{align}}

\singleboxs{Pattern concatenation, union, and conditioning:}{
\small
\begin{align}
    \llbracket \pi_1\pi_2 \rrbracket_{G} &= \{ \beta_1 \Join \beta_2 \mid \beta_i \in \llbracket \pi_i \rrbracket_{G}, i=1,2 \text{ and } \beta_1 \sim \beta_2 \} \label{Semantics:concatenation} \\
    \llbracket \pi_1 \syntaxFont{+} \pi_2 \rrbracket_{G} &= \{ \beta \cup \beta' \mid \beta \in \llbracket \pi_1 \rrbracket_{G} \cup \llbracket \pi_2 \rrbracket_{G} \} \label{Semantics:union} \\
    \llbracket \pi \ \syntaxFont{WHERE} \ \Phi \rrbracket_{G} &= \{ \beta \in \llbracket \pi \rrbracket_{G} \mid \llbracket \Phi \rrbracket^\beta_{G} = True \} \label{Semantics:Conditions}
\end{align}}
\noindent In \eqref{Semantics:union}, $\beta'$ maps any variable in ${\llbracket \pi_1 \syntaxFont{+} \pi_2 \rrbracket}_G$ not in $\beta$'s domain to $Null$.

\singleboxs{Graph patterns:}{
\begin{equation}
    \llbracket \Pi_1, \Pi_2 \rrbracket_{G} = \{ \beta_1 \Join \beta_2 \mid \beta_i \in \llbracket \Pi_i \rrbracket_{G}, i=1,2 \text{ and } \beta_1 \sim \beta_2 \} \label{Semantics:graphpatterns}
\end{equation}}

\noindent The semantics $\llbracket \expression \rrbracket^\beta_{G}$ of an expression \expression{} is computed with respect to binding $\beta$ over graph $G$. For a variable \syntaxFont{x} in $\beta$'s domain, $\llbracket \syntaxFont{x} \rrbracket^\beta_{G} = \beta(x)$.

\singleboxs{Node and property values:}{
    \small
\begin{equation}
    \llbracket \syntaxFont{x.}\pkey \rrbracket^\beta_{G} = 
                \begin{cases}
    Val(p) & \text{if } \beta(\syntaxFont{x}) \in N \cup E, p \in \sigma(\beta(\syntaxFont{x})), Key(p) = \pkey \\
    Null & \text{otherwise}
    \end{cases} \label{Semantics:nodeandpropertyvalue}
\end{equation}}
\singleboxs{Property and Label set operations:}{
\begin{align}
    &\llbracket \syntaxFont{KEY(x)} \rrbracket^\beta_{G} = Key(\llbracket \syntaxFont{x} \rrbracket^\beta_{G}) \text{ for } \llbracket \syntaxFont{x} \rrbracket^\beta_{G} \in P \label{Semantics:keymacro} \\
    &\llbracket \syntaxFont{VAL(x)} \rrbracket^\beta_{G} = Val(\llbracket \syntaxFont{x} \rrbracket^\beta_{G}) \text{ for } \llbracket \syntaxFont{x} \rrbracket^\beta_{G} \in P \label{Semantics:valmacro} \\
    &\llbracket \syntaxFont{LABEL(x)}\rrbracket^\beta_{G} = \mu(\llbracket \syntaxFont{x} \rrbracket^\beta_{G}) \text{ for } \llbracket \syntaxFont{x} \rrbracket^\beta_{G} \in \labels\label{Semantics:LABELQmacro}
\end{align}}
\noindent The semantics $\llbracket \Phi \rrbracket^\beta_{G}$ of a condition $\Phi$ is an element in $\{True, False, Null\}$, computed with respect to $\beta$ as follows.

\singleboxs{Conditions:}{
\small
\begin{align}
    \llbracket \expression_1 = \expression_2 \rrbracket^\beta_{G} &= 
                \begin{cases}
    Null & \text{if } \llbracket \expression_1 \rrbracket^\beta_{G} = Null \text{ or } \llbracket \expression_2 \rrbracket^\beta_{G} = Null \\
    True & \text{if } \llbracket \expression_1 \rrbracket^\beta_{G} = \llbracket \expression_2 \rrbracket^\beta_{G} \\
                False & \text{otherwise}
    \end{cases} \label{Semantics:equal} \\[2ex]
    \llbracket \expression_1 < \expression_2 \rrbracket^\beta_{G} &= 
                \begin{cases}
    Null & \text{if } \llbracket \expression_1 \rrbracket^\beta_{G} = Null \text{ or } \llbracket \expression_2 \rrbracket^\beta_{G} = Null \\
    True & \text{if } \llbracket \expression_1 \rrbracket^\beta_{G} < \llbracket \expression_2 \rrbracket^\beta_{G} \\
                False & \text{otherwise}
    \end{cases} \label{Semantics:lessthan} \\[2ex]
    \llbracket \syntaxFont{x:\olabel} \rrbracket^\beta_{G} &= 
                \begin{cases}
    True & \text{if } \llbracket \syntaxFont{x} \rrbracket^\beta_{G} \in N \cup E \text{ and } \olabel \in \mu(\lambda(\llbracket \syntaxFont{x} \rrbracket^\beta_{G})) \\
    False & \text{if } \llbracket \syntaxFont{x} \rrbracket^\beta_{G} \in N \cup E \text{ and } \olabel \notin \mu(\lambda(\llbracket \syntaxFont{x} \rrbracket^\beta_{G}))
    \end{cases} \label{Semantics:objectlabelcondition}
\end{align}}
\singleboxs{Logical operations:}{
\begin{align}
    \llbracket \Phi_1 \ \syntaxFont{AND} \ \Phi_2 \rrbracket^\beta_{G} &= \llbracket \Phi_1 \rrbracket^\beta_{G} \wedge \llbracket \Phi_2 \rrbracket^\beta_{G} \label{Semantics:and} \\
    \llbracket \Phi_1 \ \syntaxFont{OR} \ \Phi_2 \rrbracket^\beta_{G} &= \llbracket \Phi_1 \rrbracket^\beta_{G} \vee \llbracket \Phi_2 \rrbracket^\beta_{G} \label{Semantics:or} \\
    \llbracket \syntaxFont{NOT} \ \Phi \rrbracket^\beta_{G} &= \neg \llbracket \Phi \rrbracket^\beta_{G} \label{Semantics:not}
\end{align}}
\singleboxs{Set operation:}{
\begin{align}
    \llbracket c \ \syntaxFont{ELEMENTOF} \ \syntaxFont{y} \rrbracket^\beta_{G} &= 
                \begin{cases}
    True & c \in \mu(\llbracket \syntaxFont{y} \rrbracket^\beta_{G}) \\
                False & \text{otherwise}
    \end{cases} \label{Semantics:ELEMENTOFmacro} \\[2ex]
    \llbracket \syntaxFont{SUBSETEQ(x, y)}\rrbracket^\beta_{G} &= 
                \begin{cases}
    True & \mu(\llbracket \syntaxFont{x} \rrbracket^\beta_{G}) \subseteq \mu(\llbracket \syntaxFont{y} \rrbracket^\beta_{G}) \\
                False & \text{otherwise}
    \end{cases} \label{Semantics:SUBSETEQmacro}
\end{align}}

\begin{definition}
    \MGPML{} evaluates queries and clauses using working tables, ensuring consistency with GQL standards. A table T comprises bindings with shared domains.
\end{definition}

\noindent Finally, we define the semantics of clause \clauses{} and query \queries{} as functions operating on table T for graph $G$.
\singleboxs{Clauses and queries:}{
\begin{align}
    &\llbracket \syntaxFont{MATCH} \ \graphPatterns \rrbracket_{G} (T) = \bigcup_{\beta \in T} \{ \beta \Join \beta' \mid \beta' \in \llbracket \graphPatterns \rrbracket_{G}, \beta \sim \beta' \} \label{Semantics:matchclause} \\[2ex]
    &\llbracket \syntaxFont{FILTER} \ \conditions \rrbracket_{G} (T) = \{ \beta \in T \mid \llbracket \conditions \rrbracket^\beta_{G} = True \} \label{Semantics:filterclause} \\[2ex]
    &\llbracket \clauses \ \queries \rrbracket_{G} (T) = \llbracket \queries \rrbracket_{G} (\llbracket \clauses \rrbracket_{G} (T)) \label{Semantics:query} \\[2ex]
    &\llbracket \syntaxFont{RETURN} \ \expression_1 \ \syntaxFont{AS} \ x_1, \ldots ,\expression_n \ \syntaxFont{AS} \ x_n \rrbracket_{G} (T) = \bigcup_{\beta \in T} \{ (x_1 \mapsto \llbracket \expression_1 \rrbracket^\beta_{G}, \ldots, x_n \mapsto \llbracket \expression_n \rrbracket^\beta_{G}) \} \label{Semantics:return}
\end{align}}
\noindent The semantics of \MGPML{} build upon and extend those of GPML, providing a robust framework for querying meta-property graphs. Key features include the ability to match nodes and edges with specific labels or properties, handle meta-nodes that encapsulate subgraphs, and perform operations on property and label objects. The language supports pattern concatenation, union, and conditioning, as well as set operations like \syntaxFont{ELEMENTOF} and \syntaxFont{SUBSETEQ}. Clauses such as \syntaxFont{MATCH} and \syntaxFont{FILTER}, along with the \syntaxFont{RETURN} statement, allow for complex query construction. These foundations enable powerful and flexible querying capabilities, facilitating advanced analytics and metadata management in PG databases.

\section{Next steps towards the \MPG{} vision}
\label{sec:futureWorks}

In this paper, we highlighted the critical need to break down barriers between data and metadata in property graph management. As the first concrete steps towards this vision, we introduced a fully specified data model and query language for meta-property graphs, enabling seamless modeling and interoperation of data and metadata. While this work lays the foundation for flexible property graph management for contemporary applications, further research is needed to fully realize this vision.
    \paragraph*{\textbf{(1) Physical implementation and technical challenges.}} A key challenge is implementing the \MPG{} data model efficiently. Since label sets and properties are treated as data objects with identifiers, they may need dedicated storage and indexing strategies. Alternatively, approaches like concatenated IDs could maintain existing storage structures but may impact query evaluation. Research is needed on physical representations and indexing strategies that optimize \MPG{} performance.
    \paragraph*{\textbf{(2) Meta-Property Graphs in practice.}} We need to investigate how \MetaPG{} can enhance knowledge engineering and management in practice. Understanding how metadata awareness and substructure reification can contribute to improving tasks such as auditing and human-in-the-loop validation of knowledge graphs or data cleaning, wrangling, integration, and exchange is crucial. Furthermore, \MPGs{} can facilitate knowledge reasoning and analysis on knowledge graphs by introducing inherent subgraph annotation and querying different forms of metadata. Additionally, developing effective educational approaches and training resources for students and professionals working with \MPGs{} and \MGPML{} requires further study.
    \paragraph*{\textbf{(3) Improvements and integration.}} \MetaPG{} and \MGPML{} can be enhanced through: (1) extending \MGPML{} with additional functions to better leverage metadata awareness and existing property graph capabilities like paths, (2) developing schema and constraint languages for MPG building on PG-SCHEMA \cite{pgschema}, and (3) incorporating other forms of metadata like reflection to expand the metadata awareness in \MPG{}.

\bibliographystyle{plain}
\bibliography{ref}

\clearpage

\appendix
\section{Appendix}
\label{sec:appendixA}
This appendix provides an in-depth exploration of the semantics missing from section \ref{sub-sec:semantics}, supplementing the concise overview presented in the main body of the paper. For those interested in a more detailed understanding, the following sections offer comprehensive insights into the formal structures and rules that underpin the query language's design and functionality. This additional material aims to enhance the discussion's clarity and completeness without disrupting the main text's narrative flow. \\

    \begin{flalign}\label{Semantics:nodeWithLable}
        & \llbracket \syntaxFont{(:\olabel)} \rrbracket_{G}  =  \{ ()  \mid  n \in N , \olabel \in \mu(\lambda(n))\} &
    \end{flalign}

    \begin{flalign}\label{Semantics:meta-node}
        & \llbracket \syntaxFont{(::$\pi$)} \rrbracket_{G}  = 
         \llbracket\syntaxFont{()}\rrbracket_{G } \Join \llbracket \pi \rrbracket_{G_n} &
    \end{flalign}

    \begin{flalign}\label{Semantics:meta-nodeWithLable}
        & \llbracket \syntaxFont{(:\olabel::$\pi$)} \rrbracket_{G}  = 
         \llbracket\syntaxFont{(:\olabel)}\rrbracket_{G } \Join \llbracket \pi \rrbracket_{G_n} &
    \end{flalign}

    \begin{flalign}\label{Semantics:meta-nodeWithVariable}
            & \llbracket \syntaxFont{(x::$\pi$)} \rrbracket_{G}  = 
             \llbracket\syntaxFont{(x)}\rrbracket_{G} \Join \llbracket \pi \rrbracket_{G_n} &
    \end{flalign}

    \begin{flalign}\label{Semantics:nodeWithLableVariable}
            & \llbracket \syntaxFont{(:?y)} \rrbracket_{G}  =  \{ (\syntaxFont{y} \mapsto l)  \mid  n \in N , l = \lambda(n) \}&
    \end{flalign}

    \begin{flalign}\label{Semantics:meta-nodeWithLableVariable}
            & \llbracket \syntaxFont{(:?y::$\pi$)} \rrbracket_{G}  = 
             \llbracket\syntaxFont{(:?y)}\rrbracket_{G} \Join \llbracket \pi \rrbracket_{G_n}&
    \end{flalign}

    \begin{flalign}\label{Semantics:meta-nodeWithVariableAndLableVariable}
            & \llbracket \syntaxFont{(x:?y::$\pi$)} \rrbracket_{G}  = 
             \llbracket\syntaxFont{(x:?y)}\rrbracket_{G } \Join \llbracket \pi \rrbracket_{G_n}&
    \end{flalign}

    \begin{flalign}\label{Semantics:nodeWithPropertyVariable}
            & \llbracket \syntaxFont{().z} \rrbracket_{G}  =  \{ (\syntaxFont{z}\mapsto p)  \mid  n \in N , p \in \sigma(n) \}&
    \end{flalign}
            
    \begin{flalign}\label{Semantics:nodeWithLableAndPropertyVariable}
            & \llbracket \syntaxFont{(:\olabel).z} \rrbracket_{G}  =  \{ (\syntaxFont{z}\mapsto p)  \mid  n \in N , \olabel \in \mu(\lambda(n)), p \in \sigma(n)\}&
    \end{flalign}

    \begin{flalign}\label{Semantics:meta-nodeWithProperty}
            & \llbracket \syntaxFont{(::$\pi$).z} \rrbracket_{G}  = 
             \llbracket\syntaxFont{().z}\rrbracket_{G } \Join \llbracket \pi \rrbracket_{G_n}&
    \end{flalign}

    \begin{flalign}\label{Semantics:nodeWithLableAndNodeAndPropertyVariable}
            & \llbracket \syntaxFont{(x:\olabel).z} \rrbracket_{G}  =  \{ (\syntaxFont{x} \mapsto n, \syntaxFont{z}\mapsto p)  \mid  n \in N, \olabel \in \mu(\lambda(n)), p \in \sigma(n) \}&
    \end{flalign}

    \begin{flalign}\label{Semantics:meta-nodeWithLableAndProperty}
            & \llbracket \syntaxFont{(:\olabel::$\pi$).z} \rrbracket_{G}  = 
             \llbracket\syntaxFont{(:\olabel).z}\rrbracket_{G } \Join \llbracket \pi \rrbracket_{G_n}&
    \end{flalign}

    \begin{flalign}\label{Semantics:meta-nodeWithVariableAndProperty}
            & \llbracket \syntaxFont{(x::$\pi$).z} \rrbracket_{G}  = 
             \llbracket\syntaxFont{(x).z}\rrbracket_{G } \Join \llbracket \pi \rrbracket_{G_n}&
    \end{flalign}

    \begin{flalign}\label{Semantics:meta-nodeWithVariableAndLableAndProperty}
            & \llbracket \syntaxFont{(x:\olabel::$\pi$).z} \rrbracket_{G}  = 
             \llbracket\syntaxFont{(x:\olabel).z}\rrbracket_{G } \Join \llbracket \pi \rrbracket_{G_n}&
    \end{flalign}

    \begin{flalign}\label{Semantics:nodeWithLableVariableAndPropertyVariable}
            & \llbracket \syntaxFont{(:?y).z} \rrbracket_{G}  =  \{ (\syntaxFont{y} \mapsto l, \syntaxFont{z}\mapsto p)  \mid  n \in N , l = \lambda(n), p \in \sigma(n) \}&
    \end{flalign}

    \begin{flalign}\label{Semantics:nodeWithNodeAndLableAndPropertyVariable}
            & \llbracket \syntaxFont{(x:?y).z} \rrbracket_{G}  =  \{ (\syntaxFont{x} \mapsto n, \syntaxFont{y} \mapsto l, \syntaxFont{z}\mapsto p)  \mid  n \in N , l = \lambda(n), p \in \sigma(n) \}&
    \end{flalign}

    \begin{flalign}\label{Semantics:meta-nodeWithLableVariableAndProperty}
            & \llbracket \syntaxFont{(:?y::$\pi$).z} \rrbracket_{G}  = 
             \llbracket\syntaxFont{(:?y).z}\rrbracket_{G } \Join \llbracket \pi \rrbracket_{G_n}&
    \end{flalign}

    \begin{flalign}\label{Semantics:meta-nodeWithVariableAndLableVariableAndProperty}
            & \llbracket \syntaxFont{(x:?y::$\pi$).z} \rrbracket_{G}  = 
             \llbracket\syntaxFont{(x:?y).z}\rrbracket_{G } \Join \llbracket \pi \rrbracket_{G_n}&
    \end{flalign}

    \begin{flalign}\label{Semantics:udEdge}
            & \llbracket \syntaxFont{-\lbrack \rbrack-} \rrbracket_{G} = \{ ()  \mid  e \in E_u \}&  
    \end{flalign}
    
    \begin{flalign}\label{Semantics:udEdgeWithVariable}
            &\llbracket \syntaxFont{-\lbrack x\rbrack-} \rrbracket_{G} = \{ (\syntaxFont{x} \mapsto e)  \mid  e \in E_u \}& 
    \end{flalign}

    \begin{flalign}\label{Semantics:udEdgeWithLable}
            &\llbracket \syntaxFont{-\lbrack :\olabel \rbrack-} \rrbracket_{G} = \{ ()  \mid  e \in E_u , \olabel \in \mu(\lambda(e))\}& 
    \end{flalign}

    \begin{flalign}\label{Semantics:udEdgeWithLableVariable}
            &\llbracket \syntaxFont{-\lbrack :?y\rbrack-} \rrbracket_{G} = \{ (\syntaxFont{y} \mapsto l)  \mid  e \in E_u, l = \lambda(e)\}& 
    \end{flalign}

    \begin{flalign}\label{Semantics:udEdgeWithVariableAndLableVariable}
            &\llbracket \syntaxFont{-\lbrack x:?y\rbrack-} \rrbracket_{G} = \{ (\syntaxFont{x} \mapsto e, \syntaxFont{y} \mapsto l)  \mid  e \in E_u, l = \lambda(e)\}& 
    \end{flalign}

    \begin{flalign}\label{Semantics:udEdgeWithPropertyVariable}
            & \llbracket \syntaxFont{-\lbrack  \rbrack.z-} \rrbracket_{G} = \{ (\syntaxFont{z}\mapsto p)  \mid  e \in E_u, p \in \sigma(e) \}&  
    \end{flalign}

    \begin{flalign}\label{Semantics:udEdgeWithVariableAndPropertyVariable}
            &\llbracket \syntaxFont{-\lbrack x\rbrack.z-} \rrbracket_{G} = \{ (\syntaxFont{x} \mapsto e, \syntaxFont{z}\mapsto p)  \mid  e \in E_u, p \in \sigma(e) \}& 
    \end{flalign}

    \begin{flalign}\label{Semantics:udEdgeWithLableAndPropertyVariable}
            &\llbracket \syntaxFont{-\lbrack :\olabel \rbrack.z-} \rrbracket_{G} = \{ (\syntaxFont{z}\mapsto p)  \mid  e \in E_u , \olabel \in \mu(\lambda(e)), p \in \sigma(e)\}& 
    \end{flalign}

    \begin{flalign}\label{Semantics:udEdgeWithLableAndVAriableAndPropertyVariable}
            &\llbracket \syntaxFont{-\lbrack x:\olabel \rbrack.z-} \rrbracket_{G} = \{ (\syntaxFont{x} \mapsto e, \syntaxFont{z}\mapsto p)  \mid  e \in E_u , \olabel \in \mu(\lambda(e)), p \in \sigma(e)\}& 
    \end{flalign}

    \begin{flalign}\label{Semantics:udEdgeWithLableVariableAndPropertyVariable}
            &\llbracket \syntaxFont{-\lbrack :?y\rbrack.z-} \rrbracket_{G} = \{ (\syntaxFont{y} \mapsto l, \syntaxFont{z}\mapsto p)  \mid  e \in E_u, l = \lambda(e), p \in \sigma(e)\}& 
    \end{flalign}

    \begin{flalign}\label{Semantics:udEdgeWithVariableAndLableVariableAndPropertyVariable}
            &\llbracket \syntaxFont{-\lbrack x:?y\rbrack.z-} \rrbracket_{G} = \{ (\syntaxFont{x} \mapsto e, \syntaxFont{y} \mapsto l, \syntaxFont{z}\mapsto p)  \mid  e \in E_u, l = \lambda(e), p \in \sigma(e)\}& 
    \end{flalign}

    \begin{flalign}\label{Semantics:dEdgeWithLable}
            &\llbracket \syntaxFont{-\lbrack x:\olabel \rbrack->} \rrbracket_{G} = \{ (\syntaxFont{x} \mapsto e)  \mid  e \in E_d , \olabel \in \mu(\lambda(e))\}& 
    \end{flalign}

    \begin{flalign}\label{Semantics:dEdgeWithLableVariable}
            &\llbracket \syntaxFont{-\lbrack :?y\rbrack->} \rrbracket_{G} = \{ (\syntaxFont{y} \mapsto l)  \mid  e \in E_d, l = \lambda(e)\}& 
    \end{flalign}

    \begin{flalign}\label{Semantics:dEdgeWithPropertyVariable}
            & \llbracket \syntaxFont{-\lbrack \rbrack.z->} \rrbracket_{G} = \{ (\syntaxFont{z}\mapsto p)  \mid  e \in E_d, p \in \sigma(e) \}&  
    \end{flalign}

    \begin{flalign}\label{Semantics:dEdgeWithLableAndPropertyVariable}
            &\llbracket \syntaxFont{-\lbrack :\olabel \rbrack.z->} \rrbracket_{G} = \{ (\syntaxFont{z}\mapsto p)  \mid  e \in E_d , \olabel \in \mu(\lambda(e)), p \in \sigma(e)\}& 
    \end{flalign}

    \begin{flalign}\label{Semantics:dEdgeWithLableAndVAriableAndPropertyVariable}
            &\llbracket \syntaxFont{-\lbrack x:\olabel \rbrack.z->} \rrbracket_{G} = \{ (\syntaxFont{x} \mapsto e, \syntaxFont{z} \mapsto p)  \mid  e \in E_d , \olabel \in \mu(\lambda(e)), p \in \sigma(e)\}& 
    \end{flalign}

    \begin{flalign}\label{Semantics:dEdgeWithLableVariableAndPropertyVariable}
            &\llbracket \syntaxFont{-\lbrack :?y\rbrack.z->} \rrbracket_{G} = \{ (\syntaxFont{y} \mapsto l, \syntaxFont{z}\mapsto p)  \mid  e \in E_d, l = \lambda(e), p \in \sigma(e)\}& 
    \end{flalign}

    \begin{flalign}\label{Semantics:dEdgeWithVariableAndLableVariableAndPropertyVariable}
            &\llbracket \syntaxFont{-\lbrack x:?y\rbrack.z->} \rrbracket_{G} = \{ (\syntaxFont{x} \mapsto e, \syntaxFont{y} \mapsto l, \syntaxFont{z}\mapsto p)  \mid  e \in E_d, l = \lambda(e), p \in \sigma(e)\}& 
    \end{flalign}

\end{document}